\documentclass[twocolumn,tighten]{aastex63}

\usepackage{soul}
\usepackage{amsmath}
\usepackage[flushleft]{threeparttable}
\shorttitle{Iteration of two-dimensional slices}
\shortauthors{Jayson}
\graphicspath{{./}{figures/}}

\begin{document}

\title{Evaluation of multi-parameter likelihoods through iteration of two-dimensional slices}

\email{1jsjayson@gmail.com}

\author[0000(-)0001(-)6546(-)1850]{Joel S. Jayson}
\affil{Unaffiliated}
 \nocollaboration{1}

\begin{abstract}
  In this paper we introduce a method for resolving multi-parameter likelihoods by fixing all parameter values, but two. Evaluation of those two variables is followed by iteratively cycling through each of the parameters in turn until convergence.  We test the technique on the temperature power spectrum of the lensed cosmic microwave background (CMB).  That demonstration is particularly effective since one of the six parameters that define the power spectra, the power spectrum amplitude, $A_{s}$, nears linearity at small deviations, reducing computation to incrementation in one-dimension, rather than over a 2D grid.  At each iterative step $A_{s}$ is paired with a different parameter.  The iterative process yields parameter values in agreement with those derived by \textit{Planck}, and results are obtained within a few hundred calls for spectra. We further compute parameter values as a function of maximum multipole, $\ell_{\text{max}}$, spanning a range from $\ell_{\text{max}}$=959 to 2500, and uncover bi-modal behavior at the lower end of that range. In the general case, in which neither variable is linear, we identify moderating factors, such as changing both parameters each iterative step, reducing the number of steps per iteration.  Markov chain Monte Carlo (MCMC) computation has been the dominant instrument for evaluating multi-parameter functions.  For applications with a quasi-linear variable such as, $A_s$, the 2D iterative method is orders of magnitude more efficient than MCMC.  
\end{abstract}

\section{Introduction} \label{sec:intro}

We evaluate multi-parameter likelihoods by freezing all parameter values, with the exception of two.  Upon solution of those two parameters we step to a second pair, continuing in this fashion through all parameters, and then repeating the process until convergence. The temperature power spectrum of the cosmic microwave background (CMB) provides a ready example for demonstration. 

 The study of CMB power spectra is a mature subject \citep{hins2013,Planck2018} and provides a solid basis for comparison.  Further, of the six parameters needed to define a spectrum, one of them is the amplitude, $A_{s}$.  With no lensing $A_{s}$ is proportional to the power spectrum.  We incorporate lensing in our analysis, which introduces nonlinearity.   Nevertheless, if variations in amplitude are small, the assumption of linearity introduces little error. The presence of the quasi-linear parameter, $A_{s}$, affords  an opportunity to appreciably reduce the number of calls for power spectra computation, incrementing in one dimension, rather than over a two dimensional grid.  We introduce an algorithm to take full advantage of that opportunity.  In general we fix  $A_{s}$ at each computational step, and multiply the power spectrum by a factor, $\gamma$, which serves as the amplitude variable.  In the iterative process $A_{s}$ is updated at each succeeding step with the algorithm,
 
\begin{equation}
\label{E:algor} A_{s, k+1}=\gamma_{k} A_{s, k},
\end{equation} 
 where k is the parameter step number. As the iterative process proceeds, $\gamma$ approaches unity, and the introduced error approaches zero.  
 
Markov chain Monte Carlo (MCMC) computation is routinely used to assess cosmological parameters, and we gauge the value of the 2D iteration process with a quasi-linear parameter against that yardstick.  Other analytic methods may provide comparable results.  These include Gaussian processes \citep{ras2006} in which the distributions of derived quantities can be obtained if the random process is assumed to have a normal distribution. More specific to the problem at hand, the application of Gaussian processes to the Hessian \citep{geoga2019} provides efficient parameter evaluation.  Additionally, neural networks \citep{hay2008}, can be trained to solve multi-parameter problems \citep{Li2007}.  The Hamiltonian Monte Carlo method introduces leapfrog steps into  MCMC, significantly improving its efficiency \citep{neal2011,betan2017}, and multi-dimensional gradient descent methods offer yet other means of evaluating parameters \citep{ruder2016}.  Caution must be taken with these latter two techniques, as, repeated numerical computation of derivatives can lead to drift away from an exact solution.    
  
The evaluation of the Hubble parameter, $H_0$, is in an ambiguous state, with differing estimations between low red shift measurements \citep{freed2019,riess2022}, and CMB measurements \citep{hins2013,Planck2018}, but also with some variation within each of those subsets.  We focus on the CMB measurements herein. The two primary satellite CMB measurements were conducted by the \textit{Planck} Collaboration, and before that  the Wilkinson Microwave Anisotropy Probe (\textit{WMAP}).  \citet{Planck2018} derive a value of $H_0$=67.4$\pm$0.5 km $\text{s}^{-1}$ $\text{Mpc}^{-1}$ with a maximum multipole, $\ell_{\text{max}}$=2508. \citet{add2016}, using \textit{Planck} data with $\ell_{\text{max}}$=1000, exact a value of 69.7$\pm$1.7 km $\text{s}^{-1}$ $\text{Mpc}^{-1}$, a discrepancy that is slightly more than 1$\sigma$.  The \textit{WMAP} result with $\ell_{\text{max}}$=1200, is $H_0$=70.0$\pm$2.2 km $\text{s}^{-1}$ $\text{Mpc}^{-1}$  \citep{hins2013}.  

The Atacama Cosmology Telescope (ACT) Collaboration \citep{ACT2020,aiola2020} derived a value of $H_{0}$=67.9$\pm$1.5 km $\text{s}^{-1}$ $\text{Mpc}^{-1}$ within the multipole range of $\ell$=575-4325, and extracted a value of 67.6$\pm$1.1 km $\text{s}^{-1}$ $\text{Mpc}^{-1}$ when combining the \textit{ACT} data with that of \textit{WMAP} to provide coverage of the lower part of the multipole range.  \textit{ACT} is one of two earth bound telescopes observing CMB.  The South Pole Telescope (\textit{SPT}) is the second one.  They report a value of  $H_{0}$=68.3$\pm$1.5 km $\text{s}^{-1}$ $\text{Mpc}^{-1} $\citep{balk2023}.  When combining the \textit{SPT} data with that of \textit{Planck} they find, $H_{0}$=67.24$\pm$0.54 km $\text{s}^{-1}$ $\text{Mpc}^{-1}$, and when combining the \textit{SPT} data with that of \textit{WMAP}  a value of $H_{0}$=68.2$\pm$1.1 km $\text{s}^{-1}$ $\text{Mpc}^{-1}$.  In summary, the CMB  $H_0$ means fall in either the range $\sim$67.4-68.3, or $\sim$69.7-70.0 km $\text{s}^{-1}$ $\text{Mpc}^{-1}$, the latter range found at $\ell_{\text{max}}\leqq$1200.  A review of $H_0$ evaluations can be found at \citet{Verde2024}.  Using our iterative procedure, we evaluate the CMB parameters as a function of $\ell_{\text{max}}$, with the aim of shedding light on  $H_0$ variations within CMB evaluations..

Section~\ref{sec:2DMLE}  introduces the ln likelihood function and its partial derivatives.  We provide an example, which employs two variables, $\gamma$, and $H_0$, to illustrate the solution of the two partial derivatives..

Section~\ref{sec:iter} addresses the iterative solutions.  In Section~\ref{subsec:qlin} the quasi-linear approach, using $A_s$ in all iterative steps, is employed to derive the CMB parameters for the \textit{Planck} data with $\ell_{\text{max}}$=2500.  Four runs were conducted, each with a different set of initial values, all producing similar results with small variance, and in agreement with \textit{Planck} results \citep{Planck2018} that were obtained using the Markov chain Monte Carlo (MCMC) method \citep{metrop1953,hast1970}.  The maximum number of calls for power spectra computations for any of the runs is less than 600.  In Section~\ref{subsec:varyl} the quasi-linear analysis is applied to evaluate CMB parameters as a function of $\ell_{\text{max}}$.  We find bi-modal behavior for the \textit{Planck} results for $\ell_{\text{max}}$$\leqslant$ 1259, and an explanation  for the variant $H_0$ CMB values.
In Section~\ref{subsec:gencase} the general case, no quasi-linear parameter, is briefly discussed.  In applications where a parameter equivalent to $A_{s}$ is lacking, calls for computing the associated function are incremented over a two dimensional grid.  We identify factors that work towards reducing the number of power spectra calls in that instance.   
In Section~\ref{sec:discuss} we discuss the results, and present our conclusions.

\section{2D maximum likelihood estimate}\label{sec:2DMLE}
The \textit{Planck} Collaboration advanced the most accurate and precise CMB results to date \citep{Planck2018}, and we use their data in this study.  Though \textit{Planck} observations covered the entire sky, the data used to compute the CMB power spectra excludes the Galactic plane, and was processed by the $\textit{Planck}$ Collaboration  to provide a full sky equivalent.  At high multipoles the likelihood distribution for a full sky tends to a Gaussian \citep{Percival2006}.  The \textit{Planck} data and the computed power spectra are binned as described in \textit{Planck} 2018 results  V \citep{Planck2018V}.  We select a bin size of $\Delta \ell$ =30 spanning multipoles 30-2500. That yields 82 full size bins and a partial last bin. Multipoles 2-29 are omitted to approach a Gaussian distribution, and to remove anomalies found in those first few multipoles, including a power deficit \citep{Planck13}.  An additional simplification is the omission of polarization in the analysis.  The approximate Gaussian distribution validates using the ln likelihood function, $ln$$\mathcal{L}(p_{1}, p_{2})$ \citep{press}, 
 \begin{equation}
\label{E:lnlik}
ln\mathcal{L}(p_{1}, p_{2})=\sum_{i=2}^{{84}}\frac{(\mathcal{\hat{D}}_{i}^{TT}-\mathcal{D}_{i}^{TT}(p_{1}, p_{2}))^{2}}{\hat{\sigma}_{i}^{2}}
\end{equation}
where,
 \begin{equation}
\label{E:D} 
\mathcal{D}^{TT}_{i}(p_{1}, p_{2})=\frac{\ell_{i}(\ell_{i}+1)\mathcal{C}_{i}^{TT}(p_{1}, p_{2})}{2\pi}
\end{equation}
denotes an element of the binned computed temperature power-spectrum, entries $\mathcal{\hat{D}}_{i}^{TT}$ and ${\hat{\sigma}_{i}^{2}}$ represent the binned data points and corresponding variance, respectively, $\ell_{i}$, the weighted multipole, and $\mathcal{C}_{i}^{TT}(p_{1}, p_{2})$, the weighted multipole component, both within each bin, and $p_{1}$ and $p_{2}$ the two parameters that we choose to vary. We use the CLASS code to evaluate the temperature  power spectra \citep{class}.

Six parameters determine the power spectrum configuration, $h$ (as defined by the Hubble parameter, $H_{0}=100h$ km $\text{s}^{-1}$ $\text{Mpc}^{-1}$), $\Omega_{b}h^{2}$, the baryon fraction, $\Omega_{c}h^2$, the cold dark matter fraction,  $\tau$, the reionization optical depth, $n_s$, the spectral index, and $A_s$, the power spectrum amplitude.  Since a precise evaluation of $\tau$ depends upon polarization data we adopt the value of 0.0543 found in \citet{Planck2018}, and use that value throughout.  We will later find it convenient to substitute $\theta_{s}$, the angular scale of the first acoustic peak, for $h$.

We initially choose $h$ as one of the variables.  The power spectrum amplitude factor, $\gamma$, serves as the second variable, and is consequential for rapid computation. Lensing has been incorporated in our investigation, which negates the linear relationship between $A_{s}$ and the power spectra.  Notwithstanding, for small excursions from the nominal amplitude, linearity is assumed  with little error.   That error is eliminated when Equation~\ref{E:algor} is incorporated into the iterative process.   $A_s$ is set to a constant value within $\mathcal{D}_{i}^{TT}(h, A_{s})$ and we assume that $\mathcal{D}^{TT}_{i}(h, A_{s})=\gamma y_{i}^{TT}(h)$, with all variation of $A_{s}$ relegated to $\gamma$.   

With that assumption, we take partial derivatives of Equation~\ref{E:lnlik}) and setting them to zero find,
\begin{equation}
\label{E:deriv}
0=\sum_{i=2}^{84}\frac{(\mathcal{\hat{D}}_{i}^{TT}-\gamma y_{i}^{TT}(h))}{\hat{\sigma}_{i}^{2}}\gamma\partial y_{i}^{TT}(h)/\partial h
\end{equation}
and,
\begin{equation}
\label{E:deriv2}0=\sum_{i=2}^{84}\frac{(\mathcal{\hat{D}^{TT}}_{i}-\gamma y_{i}^{TT}(h))}{\hat{\sigma}_{i}^{2}} y_{i}^{TT}(h).
\end{equation}

 \begin{figure}
\includegraphics[width=\columnwidth]{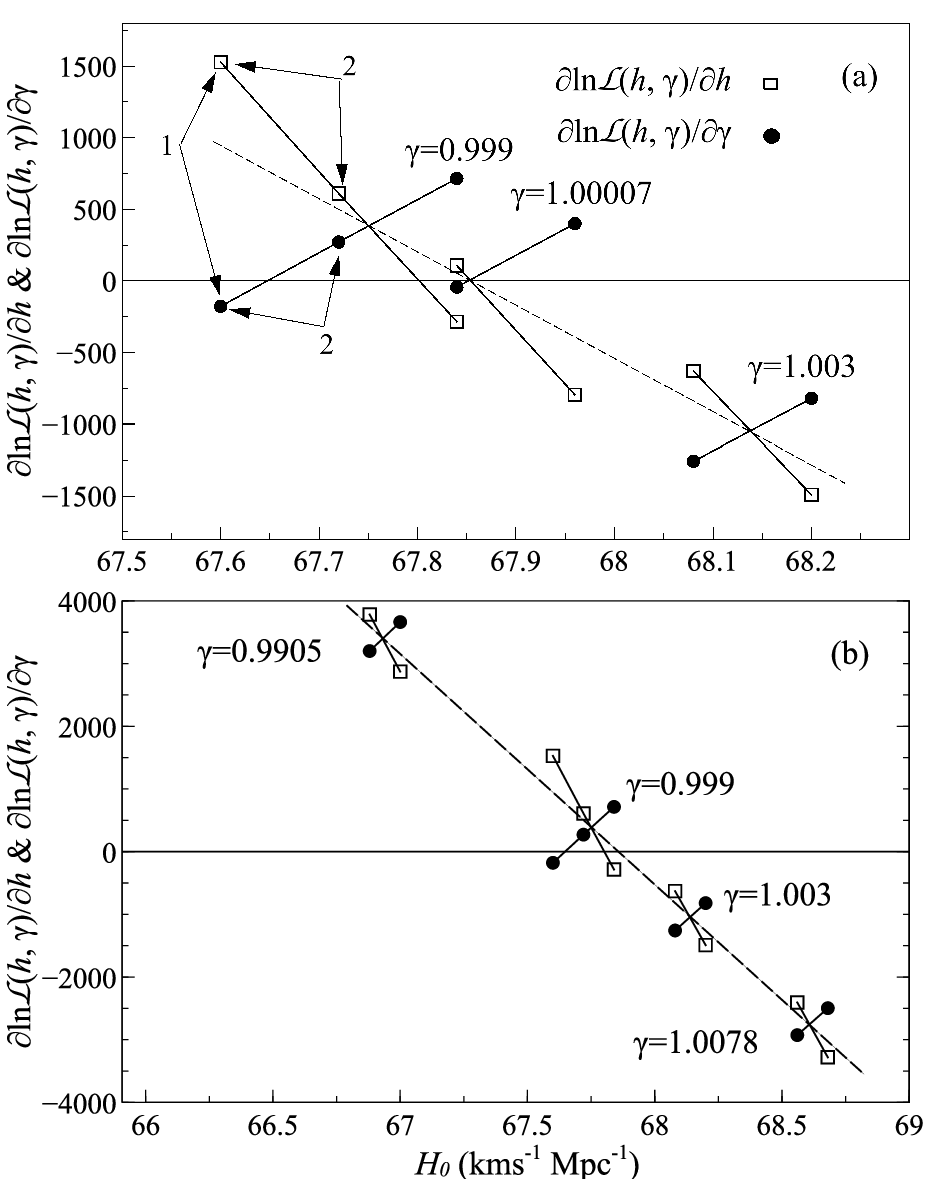}
\caption{(a) Plots of $\partial{ln\mathcal{L}(h, \gamma)}/\partial
{h}$, and $\partial{ln\mathcal{L}(h, \gamma)}/\partial{\gamma}$ versus $H_{0}$ for three pairs of crossed lines, each pair defined by a specific value of $\gamma$. The solution is found where a pair crossing lands exactly on the zero line, $\gamma$=1.00007, $h$=0.6785, in this instance. That is within 1$\sigma$ of the \textit{Planck} result.  The dashed line demonstrates that the solution can be derived directly from two crossed line pairs that straddle the zero line. The fixed parameter values are \citep{Planck2018}, $\Omega_{b}h^{2}$=0.02237, $\Omega_{c}h^2$=0.1200, $n_{s}$=0.9649, and $\tau$=0.0543. ``Pair'' of points such as those indicated with arrows and labeled, 1, are generated by the ln likelihood partial derivatives at specific values of $H_0$.  All points come in such pairs. ``Pair'' of points, such as those denoted with arrows and labeled, 2, are generated to determine line direction.   (b) Expanded scale showing that crossed line pairs with larger separations than in (a) yield the same solution.  Two crossed line pairs from (a) are shown to give a sense of scale.}
\label{fig:Fig1}
\end{figure}

Figure~\ref{fig:Fig1}a depicts the results attained using \textit{Planck} parameter values \citep{Planck2018}.  The Hubble parameter is plotted along the abscissa, while the values of the partial derivatives of the ln likelihood  are plotted along the ordinate.  

The term, ``pair''  describes three different entities that are illustrated in Figure~\ref{fig:Fig1}a. We compute two partial derivatives at a specific value along the abscissa, $H_0$ in this example.  The resulting pair of points is labeled, 1, in the figure.   When looking for which direction to continue the computation, we generate a pair of points, labeled, 2, in the figure.  The direction to proceed to generate crossed lines for a particular value of $\gamma$ is now evident.  That is the first step towards a solution.  A pair of crossed lines is our third use of the word.   Each pair of the crossed lines is defined by a specific value of $\gamma$.  Much like finding the root of a function with a single variable, once the zero line is bracketed by two pairs of crossed lines, we have the equivalent of the intermediate value theorem, and the solution on the zero line is readily found.  The solution, with $\gamma=1.00007$, $h$=06785 was derived by this bracketing and bisection technique. However, the dashed line in the figure shows that we can derive it directly from the  crossed line pairs that straddle the zero line.  The intersection of the dashed line with the zero line furnishes the value of $H_0$, and interpolation provides the $\gamma$ value of 1.00007.  The value of $h$ is higher than the value 0.6736$\pm$0.0054 found in \citet{Planck2018} but lies within  1$\sigma$ of the result.  Since we use only the TT spectra, an exact match is not expected. In Figure~\ref{fig:Fig1}b we have expanded the distance between the line pairs, finding the same solution to four significant figures.   In Section~\ref{subsec:gencase} a procedure is outlined to apply this technique in the general case to minimize the number of power spectra computation calls when neither parameter variable is linear.

\section{Iterative solutions}\label{sec:iter}
\subsection{Quasi-linear study}\label{subsec:qlin}
In Figure~\ref{fig:Fig1} we varied $h$, but any of the other parameters, i.e., $\Omega_{b}h^{2}$,  $\Omega_{c}h^2$, and $n_s$ can also be paired with $A_s$ through $\mathcal{D}^{TT}_{i}(p, A_{s})=\gamma y_{i}^{TT}(p)$.  \citet{Kos2002} introduced the use of the angular scale of the first acoustic peak, $\theta_{s}$, as one of the parameters, and subsequently that was adopted in CMB analysis as a substitute for $h$ \citep{Verde2003}. We do the same here.  The CLASS code angular scale, $\theta_{s}$, is similar, but not identical to that defined by CAMB, $\theta_{*}$ \citep{Lewis2000}.

\begin{table*}
\caption{\label{tab:table1}A sample from an iteration run that converged after 31 iterations.   Shown are the initial parameter values, the results of the first two iterations, the result of the 15th iteration, and the results of the final two iterations. The entries for $10^9$$A_{s}$/$\gamma$, such as the first one, 1.050275/1.78167, denote the value of $A_{s}$ used in computing the parameter value immediately above, and the resulting value of $\gamma$ derived from that computation. See text for further details.}
\begin{ruledtabular}
\begin{tabular}{lcccc}
&100$\theta_{s}$&$\Omega_{b}h^2$& $\Omega_{c}h^2$&$n_s$  \\
&$10^{9}A_{s}$/$\gamma$&$10^{9}A_{s}$/$\gamma$&$10^{9}A_{s}$/$\gamma$&$10^{9}A_{s}$/$\gamma$\\
\hline
Initial values&--&0.03& 0.1&1.2 \\
\hline
Iteration \#1&1.05190&0.01758&0.0708&1.1658  \\
&1.050275/1.78167&1.871243/0.9935&1.859080/0.96855&1.800611/1.0141\\
Iteration \#2&1.04372 &0.02331&0.0649&1.1246 \\
&1.826000/0.9866 &1.801532/0.99417&1.791029/0.99323&1.778903/1.0174\\
&...&...&...&...\\
Iteration \#15&1.04225&0.02251&0.1160&0.9744\\
&2.074503/0.9998& 2.074088/0.99995& 2.073984/1.00165&2.077406/1.0012\\
&...&...&...&...\\
Iteration \#30&1.04180&0.02227&0.1203&0.9649\\
&2.101274/0.999995&2.101263/0.999995&2.101253/1.00004&2.101337/0.99997\\
Iteration \#31&1.04180&0.02227&0.1203&0.9649\\
&2.101274/0.999995&2.101263/0.99999&2.101242/1.00003&2.101305/0.99998\\
\end{tabular}
\end{ruledtabular}
\end{table*}

Sequentially computing the 2D solution for each parameter, and substituting the latest value into 
$\mathcal{D}^{TT}_{i}(p, A_{s})=\gamma y_{i}^{TT}(p)$ is an iterative process. We have conducted four iterative runs, each with a different set of initial parameter values.  In two of the runs (one with 10 iterations to convergence, the other with 16)  we retain a fiducial value of $A_{s}=2.100549 \text{x} 10^{-9}$, and within specified ranges select the other parameter values at random.  For the other two runs (19 and 31 iterations), we introduce the algorithm of Equation~\ref{E:algor}. That enables a choice to arbitrarily set an initial value for  $A_{s}$.  We additionally add a requirement that $\gamma$ $\rightarrow$1 at convergence.   Table~\ref{tab:table1}  provides a sample from one of those runs (31 iterations). The parameter, 100$\theta_{s}$ is computed first, hence the lack of an initial value.  The setting for the initial value for $A_{s}$ was half that of our fiducial value (By contrast, the $A_{s}$ initial value in the second run with the algorithm was twice the fiducial value. Initial values of the other parameters were likewise chosen to be dissimilar).  The four runs yield the following results, 100$\theta_s$=1.04179$\pm$0.000026, $\Omega_{b}h^{2}$=0.02227$\pm$0.000017, $\Omega_{c}h^{2}$=0.1205$\pm$0.00026, $n_s$=0.9646$\pm$0.00057, and $10^{9}A_s$=2.102$\pm$0.0013.  We also extract values of $H_0$ from each run, and find, $H_0$=67.6$\pm$0.1 km $\text{s}^{-1}$ $\text{Mpc}$.  The variance between runs is tight. Despite our simplifications, all parameter values are in close agreement with the \textit{Planck} results \citep{Planck2018}.  

We note the following specific; the computed values of $\gamma$$A_s$ in the two runs where $A_s$ was kept at the fiducial value are in agreement with the computed values in the two runs where Equation~\ref{E:algor} was employed. For the former pair, $\gamma$ converges to a value exceeding unity by  less than 0.1\% for both runs.  For the latter pair, $\gamma$ converges to unity, but $A_s$ converges to a value slightly greater than the fiducial value. In one run $A_s$ is less than 0.1\% higher than the fiducial value, in the other, about 0.15\% higher.  

The solution derivation examples given in Figure~\ref{fig:Fig1} show crossed lines that exhibit significant orthogonality.  That follows too for the other parameters, with the exception of $n_{s}$.  The crossed lines are close to parallel in that instance.  The saving grace here is that the intersection of the crossed line pair with the zero line, is well defined.  The uncertainty lies with the derived value of $\gamma$.  Since $\gamma$ is recomputed at each parameter step, any introduced error is immediately adjusted for in the following parameter step.

 \subsection{CMB parameter values as a function of $\ell_{max}$}
 \label{subsec:varyl}

 \begin{table*}
\caption{\label{tab:table2}Compilation of CMB parameter values as a function of $\ell_{\text{max}}$. All values are computed using \textit{Planck} data, with the exception of one point computed with \textit{WMAP} data at $\ell_\text{max}$=1019, and designated 1019W. The second column denotes the number of power spectrum peaks that lie within the range delineated by $\ell_{\text{max}}$.  The peak numbers marked with a dagger, stipulate that those points lie at the summit of the indicated peak. A double dagger has been placed at $\ell_\text{max}$=1769 to signify that it is one bin shy of the apex of the 6th peak, which lies at  bin, $\ell_\text{max}$=1739.  $\ell_{\text{max}}$=1379 is labeled 4/5 since although the apex of peak \#5 has been removed, the remaining portion of that peak still affects parameter values.  See text for discussion of mode B, found at those values of $\ell_{max}$ $\leqq$ 1259.}
\begin{ruledtabular}
\begin{tabular}{lccccccc}
$\ell_{\textit{Max}}$ &\#peaks&100$\theta_{s}$&$\Omega_{b}h^2$& $\Omega_{c}h^2$&$n_s$ & $10^{9}$ $\text{A}_{\text{s}}$&\textit{h}\\
&&&&&&&0.1ln likelihood\\
\hline
959&3&1.04082$\pm$0.00072&0.02222$\pm$0.00031&0.1184$\pm$0.0028&0.9650$\pm$0.0097&2.089$\pm$0.013&0.680$\pm$0.0024\\
&&&&&&&2.1806\\
959B&3&1.04103$\pm$0.0014 &0.02232$\pm$0.00064&0.1175$\pm$0.0028&0.9682$\pm$0.01&2.085$\pm$0.013& 0.684$\pm$0.0047\\
&&&&&&&2.1848\\
1019&3&1.04083$\pm$0.00069&0.02221$\pm$0.00031&0.1185$\pm$0.0028&0.9651$\pm$0.0098&2.089$\pm$0.012&0.679$\pm$0.0023\\
&&&&&&&2.1927\\
1019B&3&1.04105$\pm$0.00070&0.02234$\pm$0.00031&0.1173$\pm$0.0027&0.9693$\pm$0.0099&2.084$\pm$0.012&0.685$\pm$0.0024\\
&&&&&&&2.2049\\
1019W&3&1.04109$\pm$0.0018&0.02238$\pm$0.00042&0.1156$\pm$0.0041&0.9648$\pm$0.011&2.070$\pm$0.029&0.692$\pm$0.0062\\
1139&4$\dagger$&1.04124$\pm$0.00048&0.02237$\pm$0.00023&0.1172$\pm$0.0020&0.9701$\pm$0.0063&2.084$\pm$0.010&0.687$\pm$0.0016\\
&&&&&&&2.5839\\
1139B &4$\dagger$&1.04125$\pm$0.00048&0.02240$\pm$0.00023&0.1170$\pm$0.0020&0.9707$\pm$0.0063&2.083$\pm$0.010&0.688$\pm$0.0016\\
&&&&&&&2.5845\\
1199&4&1.04119$\pm$0.00048&0.02230$\pm$0.00022&0.1178$\pm$0.0019&0.9679$\pm$0.0054&2.087$\pm$0.0095&0.684$\pm$0.0016\\
&&&&&&&2.6215\\
1199B&4&1.04123$\pm$0.00048&0.02234$\pm$0.00021&0.1174$\pm$0.0018&0.9690$\pm$0.0054&2.085$\pm$0.0095&0.686$\pm$0.0016\\
&&&&&&&2.6094\\
1259&4&1.04139$\pm$0.00046&0.02221$\pm$0.00021&0.1187$\pm$0.0018&0.9645$\pm$0.0053&2.091$\pm$0.0095&0.680$\pm$0.0015\\
&&&&&&&3.3284\\
1259B&4&1.04144$\pm$0.00045&0.02224$\pm$0.00021&0.1184$\pm$0.0018&0.9653$\pm$0.0053&2.090$\pm$0.0096&0.682$\pm$0.0015\\
&&&&&&&3.3358\\
1379&4/5&1.04203$\pm$0.00041&0.02244$\pm$0.00020&0.1187$\pm$0.0018&0.9688$\pm$0.0052&2.094$\pm$0.0092&0.684$\pm$0.0014\\
1439&5$\dagger$&1.04195$\pm$0.00038&0.02235$\pm$0.00018&0.1193$\pm$0.0017&0.9666$\pm$0.0046&2.097$\pm$0.0089&0.681$\pm$0.0013\\
1499&5&1.04194$\pm$0.00038&0.02220$\pm$0.00016&0.1200$\pm$0.0017&0.9636$\pm$0.0043&2.100$\pm$0.0088&0.677$\pm$0.0011\\
1769&6$\dagger$$\dagger$&1.04174$\pm$0.00033&0.02221$\pm$0.00014&0.1205$\pm$0.0016&0.9637$\pm$0.0039&2.102$\pm$0.0083&0.675$\pm$0.0011\\
2009&7$\dagger$&1.04176$\pm$0.00032&0.02228$\pm$0.00014&0.1205$\pm$0.0015&0.9646$\pm$0.0037&2.102$\pm$0.008&0.676$\pm$0.0013\\
2500&8&1.04179$\pm$0.00032&0.02227$\pm$0.00013&0.1205$\pm$0.0015&0.9646$\pm$0.0037&2.102$\pm$0.008&0.676$\pm$0.0011\\
\end{tabular}
\end{ruledtabular}
\end{table*}

Much the same as for $\ell_{max}$=2500, we compute CMB parameter values for $\ell_{max}$=959, 1019, 1139, 1199, 1259, 1379,1439,1499, 1769, and 2009. (The binning of multipoles explains the seemingly odd choice for these values.  Each represents the highest multipole of the last bin of the $\ell_{max}$ selection).  The parameter values are presented in Table~\ref{tab:table2} along with the $\ell_{max}$=2500 results.  At $\ell_{max}$=1019, in addition to the two modes obtained using \textit{Planck} data, a single mode solution for \textit{WMAP} is acquired.  The variances are derived from the diagonal terms of the inverse Fisher information matrix.  As the non-diagonal terms are all populated, that is an approximation that neglects correlations.  The Fisher information matrix  is formed by taking partial second derivatives of the ln likelihood.  That leads to sums of terms that include products of two partial derivatives, and second derivative terms multiplied by the differences of data and model multipole elements.  That latter component is negligible, since the difference terms tend to cancel at the solution. Hence, numerical computation of the second derivatives is not necessary \citep{press}.

\begin{figure}
\includegraphics[width=\columnwidth]{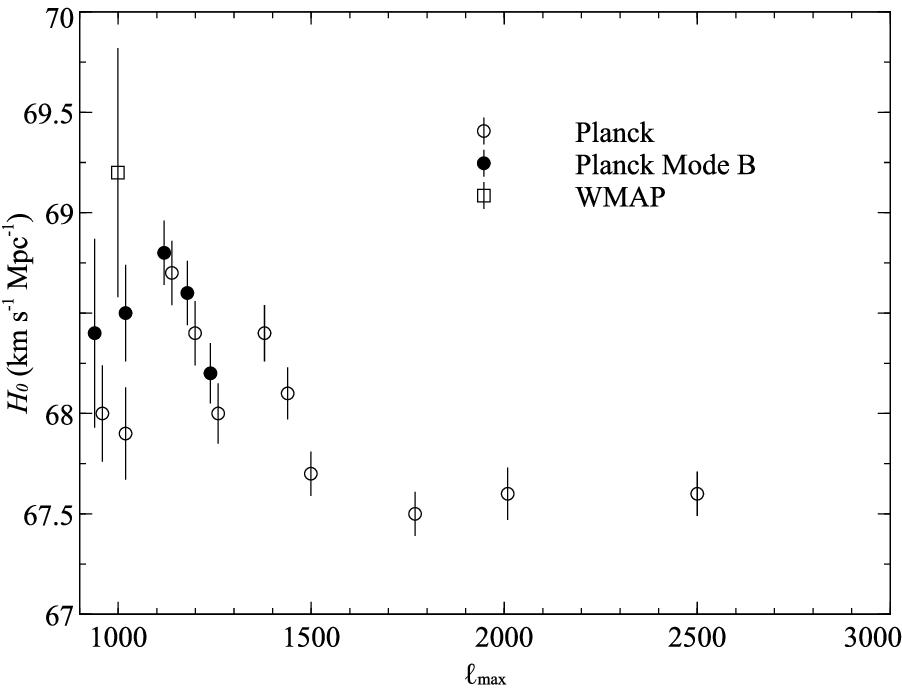}
\caption{\label{fig:Fig2}  Plot of $H_{0}$ versus $\ell_{max}$.  For those values of $\ell_{max}$ that confine no more than three, or four power spectrum peaks, i.e., 959, 1019, 1139, 1199, and 1259, a bimodal solution is obtained.  With the exception of 1019B, all B mode points, and, additionally, the 1019(\textit{WMAP}) point are shifted to the left for clarity. } 
\end{figure}

 Table~\ref{tab:table2} also lists the number of power spectrum peaks within the multipole range defined by each $\ell_{max}$.  Any change in parameter values as we decrease $\ell_{max}$ must be ascribed to a loss of information carried by the culled multipoles. $H_0$ is plotted as a function of $\ell_{\text{max}}$ in Figure~\ref{fig:Fig2}, and we use that figure to describe the results, starting at the high end, $\ell_{\text{max}}$=2500.  That range encompasses  eight power spectrum peaks.  Peak \#1 is nearly fifty times greater than peak \#8, and about twenty-five greater than peak \#7, and there is virtually no change in parameter values when peaks \#8 and \#7 are eliminated.  With the removal of peak \#6 there is a small, but noticeable change in $H_0$ at  $\ell_{max}$=1499, and that change becomes more pronounced as we cut into peak \#5.
  
 At $\ell_{\text{max}}$=1259 the effect of stripping away multipoles is substantial. The loss of peak \#5 has impacted the computed  parameter values such that there is now a major transformation.  Enough uncertainty in the solution has been introduced to support a second mode, which we refer to as mode B.   For $\ell_{\text{max}}$=1199, and 1139 the situation is similar to that of  $\ell_{\text{max}}$=1259, although $\ell_{\text{max}}$=1139 and 1139B lie at the apex of peak \#4, and the computed $H_{0}$ values reach a maximum at that point.  At the $\ell_{\text{max}}$ values of 1019 and 959 only the first three peaks remain, and the spread between the two modes has increased markedly.   The designation "parameter value" is something of a misnomer for all values of $\ell_{max}$ $\leqq$ 1259, the bimodal solution being a non-physical condition.

Turning to Table~\ref{tab:table2}, we address a couple of specifics.  First, the variances at $\ell_{max}$ =2500 are in agreement with the values obtained by the \textit{Planck} Collaboration \citep{Planck2018}, with the major exception of our $H_0$ variance, which is significantly lower than that derived by \textit{Planck}.   After attaining convergence, and computing the variances with $\theta_s$ as one of the parameters, we then solve for $h$, substitute that value of $h$ for $\theta_s$ in the CLASS program, verify that it yields the same power spectrum, and finally, recompute the variances with $h$, rather than with $\theta_s$.  In addition to finding the variance for $h$, the other variances (except for that of $\theta_s$) are recomputed along with it, with no change in their values.  That procedure instills confidence that our derived values of $h$ variances are valid.  The second particular with respect to Table~\ref{tab:table2} is the inclusion of the ln likelihood value for each mode for all $\ell_{max}$ values that exhibit a bimodal solution.  The MLE translates to a minimum in the ln likelihood (for convenience a minus sign has been removed from our definition of ln likelihood), and the global maximum is the lower of the two values.  For four of the points the base mode is at the global maximum. However, for $\ell_{max}$=1199, mode B is at the global maximum, and we are not able to reach a general conclusion.

 Where we can generalize, is with respect to the different paths to convergence taken by the two modes.  In the iterative process to obtain a solution, mode B manifests a distinct difference from the base mode in that it has a ``flat'' behavior as convergence is approached.  What we mean by ``flat'' is that even when the various parameters, 100$\theta_{s}$, $\Omega_{b}h^2$, $\Omega_{c}h^2$, $n_s$, and $10^{9}A_s$ are within a digit or two of the solution in their last significant place, several more iterations are required to reach convergence.   Strictly speaking convergence was never obtained for the $\ell_{\text{max}}$=1139, and 1199 mode B solutions.  An acutely narrow set of parameter values wandered, but never found a ``bottom''.

 \begin{table*}
\caption{\label{tab:table3}CMB parameter values computed  for the EE power spectra at $\ell_{max}$=1019.  The perceived bimodal behavior is minimal.  The relatively high CMB $h$ value is identical to that derived by \citet{Planck2018} for $\ell_{max}$=2500. } 
\begin{ruledtabular}
\begin{tabular}{lcccccc}
$\ell_{\textit{Max}}$ &100$\theta_{s}$&$\Omega_{b}h^2$& $\Omega_{c}h^2$&$n_s$ & $10^{9}$ $\text{A}_{\text{s}}$&\textit{h}\\
\hline
1019 mode 1&1.04142$\pm$0.00062&0.02331$\pm$0.00099&0.1163$\pm$0.0038&0.9646$\pm$0.012&2.093$\pm$0.019&0.699$\pm$0.0021\\
1019 mode 2&1.04142$\pm$0.00062&0.02228$\pm$0.00097&0.1164$\pm$0.0037&0.9643$\pm$0.012&2.093$\pm$0.019&0.698$\pm$0.0021\\
\end{tabular}
\end{ruledtabular}
\end{table*}

 We  briefly deviate from focusing on the TT power spectra to investigate whether or not the EE spectrum also exhibits bimodal behavior for low multipole values.  The answer as indicated in Table~\ref{tab:table3} is, just barely.  The parameter values for both modes have been locked in by convergence.  A possible interpretation is that we have captured the EE spectra in the vicinity of a branching point.  Iterative process parameter variances for TT spectra were computed in Section 3.1, on the basis of four runs at $\ell_{max}=2500$. Those variances are significantly lower than the parameter variances calculated from the inverse Fisher information matrix. Further the iterative process variances have little influence on the bi-modal results of Table 2 and Figure 2.  However, analogous iterative process parameter variances for the EE spectra offer an alternative interpretation to bi-modal behavior for the slight parameter value differences indicated in Table 3. 
 
 We are now in a position to address the variety of CMB $H_0$ results.  The discrepancy between CMB derived results and low red shift results falls outside of our scope of work.  However, we note that a recent paper \citep{kam2023}, reviews models, specifically early dark energy (EDE) models, that possibly explain the disparity between low red shift and CMB measurements.   Further refinement of CMB data is necessary before any conclusions can be reached.  
 
 There are two CMB results that require an explanation, that of \textit{WMAP}, with $\ell_{\text{max}}$=1200, $H_0$=70.0$\pm$2.2 km $\text{s}^{-1}$ $\text{Mpc}^{-1}$ \citep{hins2013}, and secondly that of \citet{add2016} using \textit{Planck} data with $\ell_{\text{max}}$=1000, 69.7$\pm$1.7 km $\text{s}^{-1}$ $\text{Mpc}^{-1}$.  The \textit{WMAP} and \textit{Planck} power spectra are in excellent agreement in this multipole range (see \citet{Planck15XI}, figure 48), except for the error level, which is significantly higher for \textit{WMAP}.  That in and of itself can explain why though our computation with the \textit{Planck} data yields two modes, that with the  \textit{WMAP} data yields one.  The difference between the two closely spaced modes is obscured in the \textit{WMAP} computation due to the higher values of $\sigma$.  Regarding the analysis in \citet{hins2013} and \citet{add2016}, there are two points to be made.  First, both studies employed MCMC to evaluate the parameters.  MCMC is deficient when evaluating multimodal solutions, and in particular when only a single mode is anticipated.  Second, both results have large variances, and our computations of $H_0$ in the relevant range $\ell_{\text{max}}$=959 to 1259 fall well within a standard deviation of either result.
 
 The main conclusion we take away concerning the dependence of CMB parameters on $\ell_{\text{max}}$, is that to arrive at an accurate solution, the spectrum should include six or more peaks.  Spectra with fewer than six peaks may serve some purpose for diagnostics, but  not for computing useful parameter values.
 
 \begin{figure}
\includegraphics[width=\columnwidth]{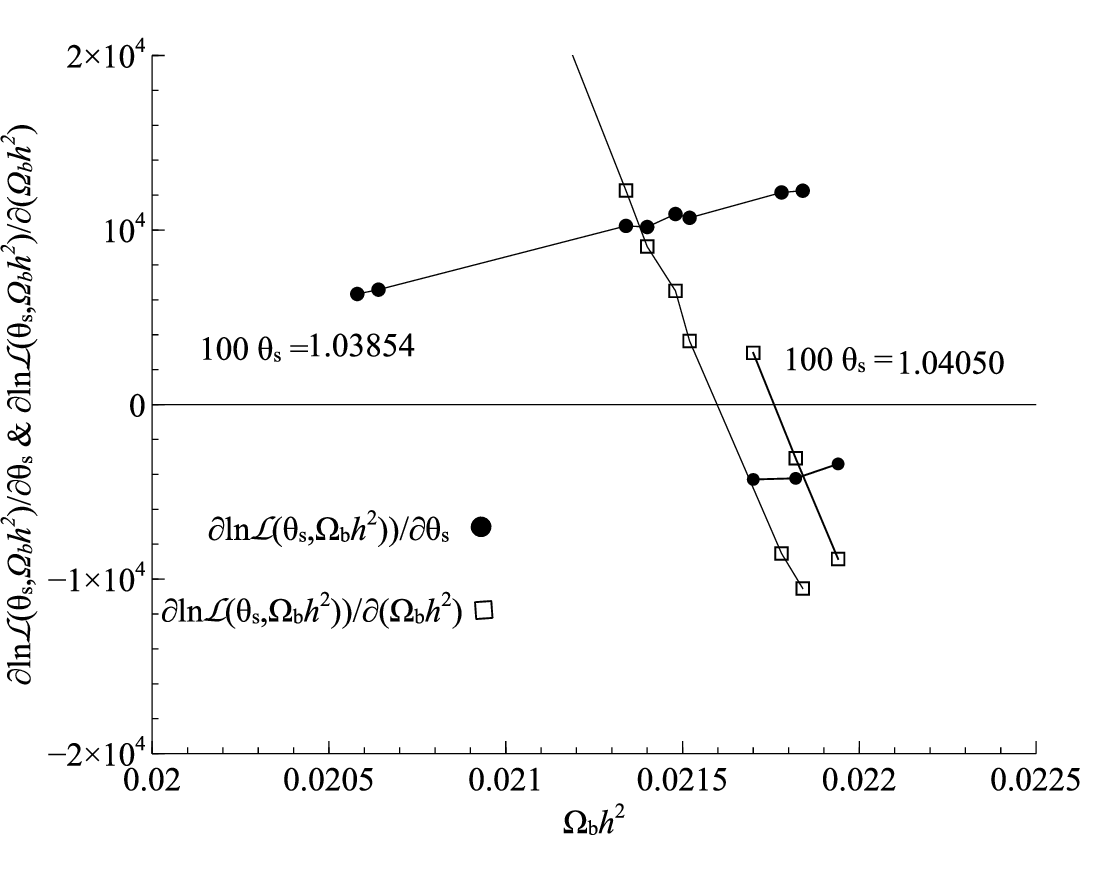}
\caption{\label{fig:Fig3} Plots of $\partial{ln\mathcal{L}(\theta_{s}, \Omega_{b}h^{2})}/\partial
{(\Omega_{b}h^{2})}$, and $\partial{ln\mathcal{L}(\theta_{s}, \Omega_{b}h^{2})}/\partial{ \theta_{s}}$ versus  $\Omega_{b}h^{2}$ for two pairs of crossed lines.  All points for both crossed line pairs require calls for power spectra computation for both  components of the ln likelihood partial derivatives. The crossed line pairs straddle the zero line, and the intercept connecting the cross point of each line pair, and  interpolation, provide an approximate solution, $\Omega_{b}h^{2}$=0.02171, and $\theta_s$=1.03994.  The connecting line is omitted in the figure for clarity.  The actual solution is $\Omega_{b}h^{2}$=0.02172, and $\theta_s$=1.04000.  Several pairs of points are plotted to demonstrate how the line crossings were located.  As depicted  in Figure~\ref{fig:Fig1}a, the ln likelihood partial derivatives generate a pair of points for each specific parameter value along the abscissa. The seeming lack of these pairs at the left is a result of the complimentary points falling beyond the figure limits. }   
\end{figure}

 \subsection{General Case}\label{subsec:gencase}
In the general case, one of the ln likelihood partials can no longer be generated by simple multiplication, and the required number of power spectra calls, or an analogous entity for a non-CMB application, is substantially increased.  Maintaining the focus on CMB, the metric by which we judge the effectiveness of a computational method is the number of calls to generate power spectra.  For the example illustrated in Table~\ref{tab:table1} fewer than 600 calls were required.  The other three runs each required 400 or fewer calls.  To tally the number of power spectra calls we must account for numerical computation of the partial derivatives.  Symmetric numerical differentiation is implemented to evaluate those partial derivatives.  For our quasi-linear approach, an isolated point requires three power spectra calls.  Generally, within a neighborhood, equally spaced points are effected to minimize the number of calls used exclusively to compute derivatives.  As demonstrated in Figure~\ref{fig:Fig1}a, when searching for the crossing of two lines, we routinely create a pair of points to determine which direction to  proceed.  That necessitates four power spectra calls in the quasi-linear case.  Figure~\ref{fig:Fig3} depicts a two parameter example in which 100$\theta_s$, and $\Omega_{b}h^{2}$ are the two variables, and we no longer have a quasi-linear condition.   The pairs of points in that figure require eight power spectra calls for each pair, as derivatives must be generated for both components of the partials.  
 
 As shown in Figure~\ref{fig:Fig1}, once the zero line is straddled by two pairs of crossed lines, the intercept of the line connecting the cross-points of each pair with the zero line identifies one parameter value to a good approximation, and interpolation identifies the other.  This procedure provides a means of reducing the number of power spectra calls.  A prudent tactic is to implement the technique, and use it to generate a value at the indicated point, putting it to the test, and rapidly closing in on a solution. (Should the zero line not be straddled by the two pairs of crossed lines, the interception of the extrapolated line joining the cross points of each pair with the zero line suffices.) This interception and interpolation (extrapolation) technique is one of three mitigating factors with regard to the required number of power spectra calls in the general case.  
 
 The second mitigating factor  pertains to  evaluating two parameters together.  In our quasi-linear example four steps were required for one iteration.  Taking two variables at once, as in Figure~\ref{fig:Fig3}, would require three steps per iteration, and in applications with larger numbers of parameters, the number of  steps needed reduces to close to half those required in the case of a linear parameter paired one at a time with the others.  
 
The third moderating factor  concerns the migration of parameter values towards their final states in a predictable manner well before final convergence.  The first two mitigating factors are particularly pertinent for the general case. This third factor is germane for both the general case, and the quasi-linear case.   Thus, in the example of Table~\ref{tab:table1}, all iterations after the fifth required less than 20 calls, that is 4-6 calls per parameter step.     This is a consequential observation, and we take a closer look.  Refer specifically to Table~\ref{tab:table1}, iteration \#15, and the $\Omega_{b}h^2$ entry, 0.02251. To obtain that result we made power spectra calls with $\Omega_{b}h^2$ values of 0.02232, and 0.02256, plus two adjoining points to compute the derivatives. Those same four $\Omega_{b}h^2$ values were used in computations through iteration \#21, at which $\Omega_{b}h^2$ was assessed as 0.02233.  For iteration \#22, having tracked the drift of  $\Omega_{b}h^2$ with increasing iteration number, the new limits for computing $\Omega_{b}h^2$ were anticipated, thus maintaining a minimum of only four power spectra calls.  At each iteration step the exact solution was obtained by tuning the value of $\gamma$ at negligible computation cost.  This example is representative of all the parameter evaluations in the quasi-linear case.   In the general case the minimum number of calls per iteration step is 24, two pairs of crossed lines, at 8 calls per pair, plus a third pair of crossed lines at the projected position to obtain an accurate solution.  In summation, this predictable trending of parameter values with increasing iteration number leads to the conclusion that the number of required power spectra calls per iteration in the quasi-linear case approaches a minimum of 4(n-1), where n is the number of parameters, and in the general case, 12n (n even) or 12(n+1) (n odd).  The question of how many iterations can be expected is dependent upon the choice of initial parameter values.  Thus, in Section~\ref{subsec:qlin}, where we were looking to establish proof of concept, initial parameter values were selected far from the known solutions, resulting in the Table~\ref{tab:table1} solution with 31 iterations.  In contrast, in Section~\ref{subsec:varyl}, where we sought to efficiently obtain solutions, for $\ell_{max}$=2009 we used the $\ell_{max}$=2500 parameter values for the initial values and found the solution in two iterations, the second iteration only needed to confirm the first.  For finding the not as easily predicted solutions for bi-modal points, we typically ran 10-15 iterations. As a result of the strong dependence upon initial parameter values, a determination of iteration number as function of the number of parameters is not readily assessed.  We suspect some dependence since a higher number of parameters increases the likelihood of outlier initial values.  Any factor so introduced  is case dependent.  Multiplying the number of iterations by the minimum number of calls per iteration confers a lower bound for the total number of power spectra calls.

 \section{Discussion}\label{sec:discuss}
 
 Within the context of multi-parameter likelihoods, two-dimensional analysis is a relatively simple procedure, made more so by the ability to visualize and plot solutions.  Making use of that facility entails setting all parameters, other than two, to fixed values and solving for those two parameters.  The full potential of that approach is realized when iteration is introduced.  All parameters are cyclically evaluated, and convergence to a solution follows.
 
 The key to regulation of computation time is the number of calls for computation of the appraised function.   For the general case, when neither variable is linear with respect to the function, calls are made over a two-dimensional grid, and the number of calls can become unwieldy.  We have identified two techniques, one regarding interpolation, the other evaluating two different parameters at each step, that reduce the number of calls, and we have also noted that the number of calls decreases sharply with increasing iteration count.  These factors keep the number of calls manageable.  However, no technique can outperform the great reduction in calls that occurs when one of the variables approaches linearity.  That is the situation with the lensed CMB power spectrum, and the result is that we are able to evaluate the cosmological parameters in good agreement with \textit{Planck}, while calling for power spectra no more than several hundred times.  That gives this 2D iterative method a decided edge over MCMC for the same computation.  For the general case, when neither variable is linear, we estimate thousands of calls per run.  We posit that in the general case, the 2D iterative method still compares favorably with MCMC computation. However, while the likelihood distribution for the CMB power spectra at high multipoles is Gaussian to a good approximation, the same is not true in general.  The matter power spectra, for example, exhibit non-Gaussian behavior \citep{tak2011}, and under those circumstances the off diagonal terms of the inverse Fisher information matrix can be significant, and lead to large errors.
 
 In our extended example of deriving cosmological parameters from power spectra we evaluate the cosmological parameters as a function of $\ell_{\text{max}}$.  The first three peaks of the CMB power spectra contain a large amount of information, and many cosmological details can be extracted from their positions and relative heights \citep{hu2001}.  Our emphasis is on how the parameter values change as multipoles are stripped from the spectra.  In that context we conclude that at least six power spectrum peaks are required to provide accurate parameter values.  Further, when the number of peaks is reduced to four, enough information has been jettisoned to lead to a bimodal solution.  That bimodal solution was not detected by \textit{WMAP}, and the higher value of $H_0$ found at low values of $\ell_{\text{max}}$ contributed to a delay in comprehending that the CMB derived value of $H_0$ was incompatible with that derived from low red shift measurements. It was a short delay, as the \textit{Planck} evaluation at $\ell_{\text{max}}$=2500 followed right on the heels of the \textit{WMAP} assessment.  
  
Though this study draws upon examples from cosmology, the 2D iterative method could find application in many areas where MCMC calculations are required to evaluate multi-parameter functions.

 \section*{acknowledgments}
 
 The author thanks the anonymous referee whose insightful comments have led to an improved manuscript.  He further acknowledges the use of the Legacy Archive for Microwave Background Data Analysis (LAMBDA), part of the High Energy Astrophysics Science Archive Center (HEASARC). HEASARC/LAMBDA is a service of the Astrophysics Science Division at the NASA Goddard Space Flight Center.


\begin{thebibliography}{99}

\bibitem[\protect\citeauthoryear{G. E. Addison et al.}{2016}]{add2016}
 Addison, G. E., Huang, Y., Watts, D. J., et al., ApJ, 818, 132 (2016)
 
 \bibitem[\protect\citeauthoryear{S. Aiola et al.}{2020}]{aiola2020}
 Aiola, S., Calabrese, E., Maurin, L., et al., JCAP, 12, 047 (2020)
 
 \bibitem[\protect\citeauthoryear{L. Balkenhol et al.}{2023}]{balk2023}
Balkenhol, L., Dutcher, D., Mancini, A.Spurio, et al., PhRvD 108, 023510 (2023)

\bibitem[\protect\citeauthoryear{M. Betancourt}{2017}]{betan2017}
Betancourt, M., arXiv:1701.02434  (2017)

\bibitem[\protect\citeauthoryear{D. Blas et al.}{2016}]{class}
 Blas, D., Lesgougues, J., \& Tram, T., JCAP, 7, 034 (2011)
 
\bibitem[\protect\citeauthoryear{S. K. Choi et al.}{2020}]{ACT2020}
  Choi, S.K., Hasselfield, M., Ho,, S.-P., et al., JCAP, 12, 045 (2020)
  
  \bibitem[\protect\citeauthoryear{W. L. Freedman et al.}{2019}]{freed2019}
  Freedman, W.L., Madore, B.F., Hatt, D., et al., ApJ, 882, 34 (2019)
  
  \bibitem[\protect\citeauthoryear{C.J. Geoga et al.}{2019}]{geoga2019}
Geoga, C.J., Anitescu, M., \& Stein, M.L., J. Comput. Graphical Stat., 29, 227 (2019)
  
\bibitem[\protect\citeauthoryear{W. K. Hastings}{1970}]{hast1970}
  Hastings, W.K., Biometrika, 57, 97 (1970)
  
 \bibitem[\protect\citeauthoryear{S. Haykin}{2008}]{hay2008}
 Haykin, S., 2008, Neural Networks and Learning Machines, (3rd ed.; New York, NY: Pearson)
  
 \bibitem[\protect\citeauthoryear{G. Hinshaw et al.}{2013}]{hins2013}
 Hinshaw, G., Komatsu, E., Spergel, D.N., et al., ApJS, 208, 19 (2013)
 
 \bibitem[\protect\citeauthoryear{W. Hu et al.}{2001}]{hu2001}
 Hu, W., Fukugita, M., Zaldarriaga, M., \& Tegmark, M., ApJ, 549, 669 (2001)
 
\bibitem[\protect\citeauthoryear{M. Kamionkowski \& A. G. Riess}{2023}]{kam2023}
Kamionkowski, M., \& Riess, A.G., ARNPS, 73, 153 (2023)

\bibitem[\protect\citeauthoryear{A. Kosowsky et al.}{2002}]{Kos2002}
Kosowsky, A., Milosavljevic, M., \& Jimenez, R., PhRvD 66, 063007 (2002)

\bibitem[\protect\citeauthoryear{A. Lewis et al.}{2000}]{Lewis2000}
 Lewis, A., Challinor, A., \& Lasenby, A., ApJ, 538, 473 (2000)

\bibitem[\protect\citeauthoryear{L. Li et al.}{2007}]{Li2007}
 Li, L.-L., Zhang, Y.-X., Zhao, Y.-H., \& Yang, D-W., CHJAA, 7, 448 (2007)

\bibitem[\protect\citeauthoryear{N. Metropolis, et al.}{1953}]{metrop1953}
Metropolis, N., Rosenbluth, A.W., Rosenbluth, M.N., et al., JChPh,, 21, 1087 (1953)

\bibitem[\protect\citeauthoryear{R.M. Neal.}{2011}]{neal2011}
Neal, R. M, (2011) In Handbook of Markov Chain Monte Carlo, (S. Brooks, A. Gelman, G. L. Jones and X.-L. Meng, eds.), CRC Press, NY 

\bibitem[\protect\citeauthoryear{W. J. Percival \& M. L. Brown}{2006}]{Percival2006}
 Percival, W. J. \&  Brown, M. L., MNRAS, 372, 1104 (2006)
  
\bibitem[\protect\citeauthoryear{\textit{Planck} Collaboration}{2014a}]{Planck13}
 \textit{Planck} Collaboration, A\&A  571, A1 (2014a) 

\bibitem[\protect\citeauthoryear{\textit{Planck} Collaboration}{2014b}]{Planck13XVI}
\textit{Planck} Collaboration, A\&A, 571, A16 (2014b) 

\bibitem[\protect\citeauthoryear{\textit{Planck} Collaboration}{2016}]{Planck15XI}
\textit{Planck} Collaboration, A\&A, 594, A11 (2016)

\bibitem[\protect\citeauthoryear{\textit{Planck} Collaboration}{2020a}]{Planck2018V}
\textit{Planck} Collaboration, A\&A, 641, A5 (2020a)  

\bibitem[\protect\citeauthoryear{\textit{Planck} Collaboration}{2020b}]{Planck2018}
 \textit{Planck} Collaboration, A\&A, 641, A6 (2020b)
 
\bibitem[\protect\citeauthoryear{W. H. Press et al.}{2007}]{press}
 Press, W.H., Teukolsky, S.A.,Vetterling, W.T., \& Flannery, B.P., 2007, Numerical Recipes, (3rd ed.; New York, NY: Cambridge Univ. Press)
 
 \bibitem[\protect\citeauthoryear{C.E Rasmussen \& C.K.I. Williams}{2006}]{ras2006}
 Rasmussen, C.E., \& Williams, C.K.I.,  2006, Gaussian Processes for Machine Learning, (MIT Press)
 
 \bibitem[\protect\citeauthoryear{A. G. Riess et al.}{2022}]{riess2022}
 Riess, A.G., Yuan, W., Macri, L.M., et al., ApJL, 934, L7 (2022)
 
  \bibitem[\protect\citeauthoryear{S. Ruder.}{2016}]{ruder2016}
 Ruder, S.,  arXiv:1609.04747 (2016)
 
 \bibitem[\protect\citeauthoryear{R. Takahashi et al.}{2011}]{tak2011}
 Takahashi, R., Yoshida, N., Takada, M., et al., ApJ, 726, 7 (2011)
   
 \bibitem[\protect\citeauthoryear{L. Verde et al.}{2003}]{Verde2003}
 Verde, L., Peiris, H. V., Spergel, D. N., et al., ApJS, 148, 195 (2003) 
 
 \bibitem[\protect\citeauthoryear{L. Verde et al.}{2024}]{Verde2024}
 Verde, L., Schöneberg, N., \& Gil-Marín, H., ARA\&A, 62, 287 (2024)  
 
\end{thebibliography}
\end{document}